\documentclass[12pt,preprint]{aastex}

\usepackage{graphicx}

\newcommand{\beq}{\begin{equation}}
\newcommand{\eeq}{\end{equation}}
\newcommand{\beqa}{\begin{eqnarray}}
\newcommand{\eeqa}{\end{eqnarray}}

\def\la{\lower.5ex\hbox{$\; \buildrel < \over \sim \;$}}
\def\ga{\lower.5ex\hbox{$\; \buildrel > \over \sim \;$}}

\begin{document}

\title{The Variety of Solutions for Dynamics in the Local Group}

\author{P.~J.~E. Peebles}  
\affil{Joseph Henry Laboratories, Princeton University, Princeton, NJ 08544}
\author{R. Brent Tully}
\affil{Institute for Astronomy, University of Hawaii, 2680 Woodlawn Drive, Honolulu, HI 96822}

\begin{abstract}
This exploration of solutions for the orbits of Local Group galaxies under the cosmological initial condition of growing peculiar velocities and fitted to measured distances, redshifts, and proper motions reveals a considerable variety of histories allowed by present constraints. The solutions also point to computations and measurements at the current level of precision that may lead to a more accurate picture of Local Group dynamics, or perhaps point to adjustments of our simple picture of the arrangement of mass within the Local Group. 
\end{abstract}
\maketitle

\section{Introduction}\label{sec:1}

Advances in measurements of galaxy distances (Jacobs, Rizzi, Tully,  et al. 2009; Dalcanton, Williams, Seth, et al. 2009; Conn, Ibata, Lewis et al. 2012) and proper motions  (Brunthaler, Reid, Falcke, et al. 2005, 2007; Sohn, Anderson,  \& van der Marel 2012; Sohn, Besla, van der Marel, et al. 2012; van der Marel, Fardal, Besla, et al. 2012; Kallivayalil, van der Marel,  Besla,  et al. 2013) are considerably tightening the constraints on dynamical models for the evolution of the galaxy distribution. We report an exploration of the variety of solutions to the dynamics of  Local Group (hereafter LG) galaxies allowed by these measurements under the assumptions that (1) LG galaxies are useful tracers of the mass concentrations at low redshift and of the evolution of the more nearly uniform mass distribution at high redshift, (2) LG galaxy peculiar velocities were small and growing at high redshift, as suggested by the gravitational instability theory for the growth of cosmic structure, and (3) the effect on dynamics within LG by external mass may be adequately represented by a few external actors. Without some variant of assumption (2) a computation of orbits back in time from measured present positions and velocities generally produces an unrealistic divergence of peculiar velocities (as $v\propto 1+z \propto  a(t)^{-1}$, where $z$ is redshift and $a(t)$ is the expansion parameter). This and other considerations motivating the dynamical model and  the Numerical Action Method  (NAM) for its application are developed and discussed in earlier papers in this series (including Peebles 1989; Peebles 1995; Peebles, Phelps, Shaya, \& Tully 2001; Peebles 2010; Peebles, Tully,  \&  Shaya 2011). We present here an exploration of model solutions allowed by LG data. Future papers in this series are intended to  analyze the large and growing fund of data on the distribution and motion of more distant as well as LG galaxies.

Section 2.1 explains the choice of the 85 model parameters, including catalog or nominal central values and uncertainties for the dynamical actor distances, redshifts, masses, proper motions, and initial peculiar velocities, and the Milky Way (MW) circular velocity. The measure of fit of solutions to the parameters is explained in Section~2.2. This measure has the familiar $\chi^2$ form, but it should be understood that the familiar statistical meaning of $\chi^2$ is inappropriate here because the dynamical model and the quantities that play the role of standard deviations are subject to considerable uncertainties. The numerical checks in Section~2.3 show that the model is computationally well defined and in reasonable agreement with the intended description of the growth of departures from homogeneity at high redshift. Section~3 presents 31 solutions at or near different local minima of $\chi^2$ allowed by generous but arguably sensible error estimates. The solutions show some degree of convergence, along with considerable diversity. These results, next steps in improving the model, and measurements that might bring acceptable solutions to better order are discussed in Section~4.

\begin{table}[htpb]
\centering
\begin{tabular}{lrrcrrrrrr}
\multicolumn{10}{c}{Table 1: The Dynamical Actors}\\
\noalign{\medskip}
\tableline\tableline\noalign{\smallskip}
 Name  & $l\ \ $ & $b\ \ $ & $D_{\rm cat}$ & $D_{\rm mod}$ & $cz_{\rm cat}$ &  $cz_{\rm mod}$
    & $v_i$ & $m_{\rm cat}$\ \ \ \  & $m_{\rm mod}$\ \ \ \  \\
   \noalign{\smallskip}
\tableline
\noalign{\smallskip}
 MW    &     $-\ \  $ &   $-\ \  $ & $ 0.0085 $ &  0.008 &  -11 &  -11 &   59 &  2.25E+01 &  1.71E+01\\
 M31   &   121.2&   -21.6& $ 0.77\pm  0.04$ &  0.817 & -301 & -306 &   61 &  2.51E+01 &  1.47E+01\\
 LMC   &   280.5&   -32.9& $ 0.049\pm  0.006$ &  0.060 &  271 &  251 &   65 &  1.24E+00 &  1.47E+00\\
 M33   &   133.6&   -31.3& $ 0.91\pm  0.05$ &  0.786 & -180 & -178 &   58 &  1.97E+00 &  3.73E+00\\
 I10   &   119.0&    -3.3& $ 0.79\pm  0.04$ &  0.899 & -348 & -340 &  120 &  4.34E-01 &  4.08E-01\\
 N185  &   120.8&   -14.5& $ 0.64\pm  0.03$ &  0.585 & -227 & -213 &   96 &  1.06E-01 &  9.58E-02\\
 N147  &   119.8&   -14.3& $ 0.73\pm  0.04$ &  0.728 & -193 & -204 &   99 &  7.65E-02 &  7.04E-02\\
 N6822 &    25.3&   -18.4& $ 0.51\pm  0.03$ &  0.575 &  -57 &  -49 &   75 &  5.95E-02 &  6.33E-02\\
 LeoI  &   226.0&    49.1& $ 0.26\pm  0.01$ &  0.264 &  284 &  294 &   57 &  2.29E-04 &  2.29E-04\\
 LeoT  &   214.9&    43.7& $ 0.41\pm  0.02$ &  0.412 &   35 &   30 &   80 &  2.69E-04 &  2.68E-04\\
 Phx   &   272.2&   -68.9& $ 0.41\pm  0.02$ &  0.407 &  -13 &   -9 &   54 &  7.35E-05 &  7.35E-05\\
 LGS3  &   126.8&   -40.9& $ 0.65\pm  0.13$ &  0.734 & -281 & -280 &   56 &  3.80E-06 &  3.80E-06\\
 CetdSp&   101.4&   -72.9& $ 0.73\pm  0.04$ &  0.728 &  -87 &  -84 &   53 &  7.85E-05 &  7.85E-05\\
 LeoA  &   196.9&    52.4& $ 0.74\pm  0.11$ &  0.598 &   28 &   34 &   79 &  3.94E-04 &  3.94E-04\\
 I1613 &   129.8&   -60.6& $ 0.75\pm  0.04$ &  0.783 & -238 & -224 &   56 &  2.85E-02 &  2.91E-02\\
 Sclp  &    68.1&   -88.3& $ 3.58\pm  0.50$ &  3.409 &  242 &  261 &   29 &  5.06E+01 &  1.23E+01\\
 Maff  &   136.4&     1.0& $ 3.61\pm  0.50$ &  4.606 &    9 &  -48 &   25 &  7.02E+01 &  6.54E+01\\
 M81   &   141.7&    41.3& $ 3.66\pm  0.50$ &  1.711 &   46 & -188 &   18 &  7.06E+01 &  1.92E+02\\
 Cen   &   311.1&    18.7& $ 3.95\pm  0.50$ &  3.882 &  518 &  528 &   18 &  1.19E+02 &  3.87E+01\\
 \noalign{\smallskip}
\tableline
\noalign{\smallskip}
\multicolumn{8}{l}{units: Mpc, km s$^{-1}$, $10^{11}m_\odot$} \\
\end{tabular}
\end{table}
\section{The Model}

\subsection{Actors and Parameters}


Table 1 lists the dynamical actors. The first eight entries are the more massive inner LG members  (assuming light is a useful measure of mass). The Small Magellanic Cloud is not included because its path relative to the Large Magellanic Cloud (LMC) may be complicated, and in any case its luminosity suggests a small mass. Dwarf galaxies that seem likely to have completed several orbits around MW or M31 are not treated as separate mass tracers; they are taken to be parts of the masses concentrated around these two large galaxies. The next seven entries in the table are the more isolated dwarfs closest to MW, out to IC\,1613 (with convenient abbreviations of names, as I1631). The masses of these dwarf galaxies are taken into account in the dynamics, but in the solutions the masses are so small that these galaxies  serve in effect as test particles. (I1613 is far enough from the other actors that its relatively large nominal mass is not very important; it too approximates a test particle.) The computation could have included a larger number of the outer LG members, but that is not done in order to allow exploration of a larger number of solutions.   

The last four actors in Table 1 are meant to approximate the effect of external mass on dynamics within LG (with Sclp representing the galaxies in and around the Sculptor Group; Maff, the galaxies in and around the Maffei-IC\,342 complex; M81,  Messier~81 and the galaxies around it; and Cen, the galaxies in and around the Centaurus group). The nominal catalog distances, angular positions and redshifts of these actors are luminosity-weighted means for the galaxies assigned to the actors. The external actor parameters are assigned generous uncertainties, as discussed in Section~2.2, because the solutions are supposed to arrive at illustrations of how dynamics within the LG may indicate the nature of the external mass distribution. 

The catalog galactic coordinates, distances, $D_{\rm cat}$, and redshifts, $cz_{\rm cat}$, in Table~1 are taken from the Local Universe (LU) catalog of distances maintained and provided on-line by Tully\footnote{http://edd.ifa.hawaii.edu select Local Universe (LU) catalog}.  The nominal masses are based on the LU tabulation of K-band luminosities with mass-to-light ratio
\beq
M/L_K=50 \hbox{  Solar units}.
\eeq
This approximates the masses found by trial fits to measured redshifts and distances. The masses of the external actors are the sums of the K-band luminosities multiplied by the same $M/L_K$. The table also lists an example of model values (with subscript mod) in one of the solutions.  

The mass distributions within LG galaxies at redshift $z<3$ are modeled as rigid spheres that produce gravitational acceleration of galaxy $i$ by galaxy $j$ at separation $r$ of the form 
\beq
g_i = Gm_j {r (r_{\rm out}^2 + r_{\rm in}^2) \over r_{\rm out}^3(r^2 +r_{\rm in}^2 )}, \qquad r < r_{\rm out},
\eeq
with the usual inverse square law at $r>r_{\rm out}$. All lengths are physical. The inner cutoff is $r_{\rm in}=10$~kpc for all galaxies. The outer cutoff for the accelerations of MW with any galaxy (apart from M31, which always is separated enough for use of the inverse square law) is computed from the MW mass and circular velocity $v_c$, both of which are parameters to be adjusted, as $r_{\rm out} = Gm_{\rm MW}/v_c^2$. The same form is computed for the acceleration of  M31 and any other galaxy from the M31 mass, which is a parameter to be adjusted, and the M31 circular velocity, which is  fixed at 250~km~s$^{-1}$. The accelerations of all other pairs of galaxies use $r_{\rm out}=100$~kpc. This simple prescription ignores the complications of overlapping halos and dynamical drag, but the approximation seems appropriate for the generally widely space orbits described below.  At $z>3$, where the actors must avoid each other so as to produce small peculiar accelerations consistent with small early peculiar velocities, the model uses the pure inverse square law.  

\begin{table}[htpb]
\centering
\begin{tabular}{lcc}
\multicolumn{3}{c}{Table 2: Proper Motions}\\
\noalign{\medskip}
\tableline\tableline\noalign{\smallskip}
  & $\mu_\alpha$  &  $\mu_\delta$  \\
\tableline
 \noalign{\smallskip}
 M31   & $    0.044\pm    0.013$ & $   -0.032\pm    0.012$ \\
 LMC   & $    1.910\pm    0.020$ & $    0.229\pm    0.047$ \\
 M33   & $    0.023\pm    0.006$ & $    0.002\pm    0.007$ \\
 I10   & $   -0.002\pm    0.008$ & $    0.020\pm    0.008$ \\
 LeoI  & $   -0.1140\pm    0.0295$ & $   -0.1256\pm    0.0293$ \\
\tableline
 \noalign{\smallskip}
\multicolumn{3}{l}{units: milli arc sec y$^{-1}$}\\
\end{tabular}
\end{table}

Table 2 lists proper motion measurements and their uncertainties, where $\mu_\alpha$ is the motion in the direction of increasing right ascension and $\mu_\delta$ is the motion in the direction of increasing declination. The proper motion of M31 is from the Hubble Space Telescope observations  (Sohn, Anderson  \& van der Marel 2012).  (The proper motion of M31 inferred from the redshifts and proper motions of other LG members by van der Marel et al. 2012 is not appropriate here because the data on other galaxies are used as separate constraints on solutions.)  The proper motion of LMC is from Kallivayalil et al. (2013), the proper motion of M33 from Brunthaler et al. (2005), the proper motion of IC\,10 (I10) is from Brunthaler et al. (2007), and the motion of LeoI from Sohn, Besla, van der Marel, et al. (2012).

Conversion from heliocentric to galactocentric velocities requires the circular velocity $v_c$ of MW at the Solar circle. The adopted value, 
\beq
v_c= 230\pm 10\hbox{ km s}^{-1}, \label{eq:vc}
\eeq
is somewhat larger than standard, in the direction indicated by Reid, Menten, Zheng, et al. (2009), and the nominal uncertainty is meant to be generous enough to allow $v_c$ to float as indicated by the dynamics. The Solar velocity relative the MW circular motion $v_c$ is taken to be $U=11.1$, $V=12.2$, $W=7.2$ km~s$^{-1}$ (Sch{\"o}nrich, Binney \& Dehnen  2010), with no allowance for uncertainty in this relatively small term. 

The model assumes the standard $\Lambda$CDM cosmology with Hubble parameter $H_o=70\hbox{ km s}^{-1}\hbox{ Mpc}^{-1}$, matter density parameter $\Omega_m=0.27$, and cosmological constant $\Omega_\Lambda =1-\Omega_m$.  The mass density in radiation is neglected. 

\subsection{Fitting Solutions to Data}

The measure of fit of a solution to the constraints is the sum of squares of differences of model and catalog values relative to the measured or nominal uncertainties, in the form 
\beq
\chi^2 = \sum_i\left({\rm model}_i - {\rm cat}_i\over  {\rm uncertainty}_i\right)^2.
\label{eq:chis}
\eeq
Though this notation is standard there are three reasons why one should not compare the value of the sum to a standard $\chi^2$ statistic. First, the dynamical model is at best a useful approximation to the  properties of  LG, so one need not expect it to produce a statistically probable fit to the measurements even if the uncertainties in the measurements were reliably known. Second, many of the uncertainties appearing in this sum are poorly known at best, and in other cases little better than guesses. Third, we must bear in mind the possibility that some of these difficult measurements been disturbed by the unanticipated systematic errors to which measurements as well as models are prey in astronomy. We do of course insist that an acceptable solution has no terms in equation~(\ref{eq:chis}) that might be judged to be implausibly large even when there are only plausible estimates of the uncertainties.

The contributions to $\chi^2$ from proper motions use the measurements and stated uncertainties in Table~2.  The MW heliocentric distance and redshift are taken to be given without uncertainty. The $\chi^2$ sum uses the catalog distances and their stated uncertainties in LU for the 14\,LG galaxies apart from MW. The sum assigns a nominal redshift uncertainty of 10~km~s$^{-1}$ to the catalog redshifts of the 14\,LG galaxies. This is not intended as a measurement uncertainty, but rather an allowance for possible motion of the visible part of a galaxy relative to the mean of its much more extended dark matter halo. The large sizes of massive halos also suggest the possibility of an offset between the distance of a galaxy and the effective distance of its dark matter halo. This is not taken into account in $\chi^2$ because there already is a significant distance measurement uncertainty. It would be logical to allow an offset of the angular position of the effective center of a dark matter halo from the angular position of its galaxy, as in Peebles, Tully, \& Shaya (2011), but for simplicity this analysis places the center of each dark matter concentration in the same direction in the sky as the galaxy.

The four external actors are allowed 50~km~s$^{-1}$ redshift uncertainty, with the intention that this and the generous allowance in distances entered in Table~1 permit the LG dynamics to point to how the actors ought to be placed. It would be very sensible to allow the angular positions of the actors to float, but again for simplicity the angular positions are fixed. 

The contributions to $\chi^2$ by the masses are 
\beq
\Delta\chi^2 _{\rm LG}= \left(\log m_{\rm model}/m_{\rm cat})\over\log {\rm 1.5} \right)^2, \qquad
\Delta\chi^2 _{\rm ext} = \left(\log m_{\rm model}/m_{\rm cat})\over\log {\rm 2} \right)^2,
\label{eq:mass-chis}
\eeq
for LG galaxies apart from MW and for the external actors. The nominal uncertainties are guesses. It is usually thought that M31 is not less massive than MW, so the contribution to $\chi^2$ from the MW mass is set at 
\beq
\Delta\chi^2_{\rm MW} = \left(\log m_{\rm MW}/m_{\rm M31})\over\log 1.1 \right)^2.
\label{eq:mass-MW}
\eeq
The more generous uncertainties in masses apart from MW allow the masses to float to what is indicated by the dynamics, while equation~(\ref {eq:mass-MW}) preserves similar MW and M31 masses, which seems reasonable. 

Peculiar velocities in the early universe are supposed to be growing as structure forms. This generally follows from the NAM numerical method, but the method can arrive at implausibly large velocities. To avoid this a galaxy is assigned contribution $f_v^2$ to $\chi^2$ when its three-dimensional peculiar velocity is $50f_v$~km~s$^{-1}$ at the starting redshift of the computation, $1+z_i=10$. The choice of 50~km~s$^{-1}$ is about what might be expected from the slow growth of peculiar velocities to several hundred kilometers per second at the present epoch. (If the initial velocity components have Gaussian distributions each galaxy velocity contributes $\chi^2$ for three degrees of freedom, but this refinement is ignored.)

The search for acceptable orbits relaxes all the model parameters to a local minimum of $\chi^2$ summed over external as well as LG parameters. However, the nature of the fit to the LG data is found to be little related to the fit to the 16 external parameters, so we take note of the fit to external parameters but do not use it as a criterion for selection of allowed solutions. 

We have explained why we do not expect a statically probable value of $\chi^2$. In line with this philosophy we adopt generous criteria for a credibly acceptable solution. No term in the $\chi^2$ sum over the 69 LG parameters may exceed four times the nominal uncertainty, and no more than 5 of the 69 differences may be between three and four times the uncertainty. Section~3 presents 31 solutions consistent with these conditions. The $\chi^2$ sums typically are about twice the number of LG parameters.  

\subsection{Computation}

The numerical methods are described in Peebles, Tully \& Shaya (2011) and earlier references therein. The search for acceptable solutions commences with random choices of distances and masses and the MW circular velocity $v_c$, within their allowed ranges, and random choices of trial orbits that are relaxed to a solution to the equations of motion by the Numerical Action Method (NAM) and the solution then relaxed to a minimum of $\chi^2$. A random trial position at each time step for each position component of an actor relative to the chosen present position, $x_o$, is of the form $x=x_o + j_1a + j_2 a(1-a)$, where the expansion parameter is $a_o=1$ at the present epoch, $j_1$ is drawn from a uniform random distribution in the interval $\pm 3$~comoving Mpc, and $j_2$ is drawn from a uniform random distribution in the interval $\pm 10$~Mpc.  Trials with positions at each time step randomly chosen from a uniform distribution in the interval $\pm 2.5$~Mpc gave similar results but produced plausible solutions at a still slower rate. Most of the solutions were found by starting with trial parameters and orbits for the first five LG galaxies in Table~1 and the four external actors. If the solution relaxed to a minimum of $\chi^2$ that looked promising the other galaxies were added one at a time. Each addition started with a random orbit and parameters that were relaxed to a solution for all the actors included so far, and the solution relaxed  to a minimum of $\chi^2$. The attempt was abandoned if the addition of a galaxy made it really difficult to fit the constraints. 

In some of the solutions the masses of the galaxies LeoI through I1613 in Table~1 were assigned starting values from the nominally allowed distribution, but apart from I1613 these masses proved to have negligible effect on the solution, the reduction of $\chi^2$ simply drawing the masses to their catalog values. Other solutions therefore were obtained starting from catalog values of these masses. 

The NAM computation uses 500 time steps uniformly spaced in the expansion parameter $a(t)$ from the starting epoch, $a_i=0.1$ (redshift $1+z_i=10$). Relaxation of the orbits to a stationary point of the action produces a solution to the equations of motion in leapfrog approximation to near machine double precision accuracy. The accuracy of each solution is checked by numerical integration of the equations of motion forward in time using 5000 steps uniformly spaced in $a$. Here the initial positions are the means of the NAM positions at the first two time steps starting from $1+z_i=10$, and the initial velocities are computed from the differences of positions at the first two NAM steps  under the assumption that the coordinate positions are varying in proportion to the expansion parameter $a(t)$, which is a good approximation. The integration forward in time, with 10 times the time steps of NAM, produces present positions and velocities that agree with the NAM  solution at $z=0$ to better than 1\,kpc and 1\,km\,s$^{-1}$. That is, there is negligible numerical error in the solutions. 

\begin{figure}[htpb]
\begin{center}
\includegraphics[angle=0,width=4.in]{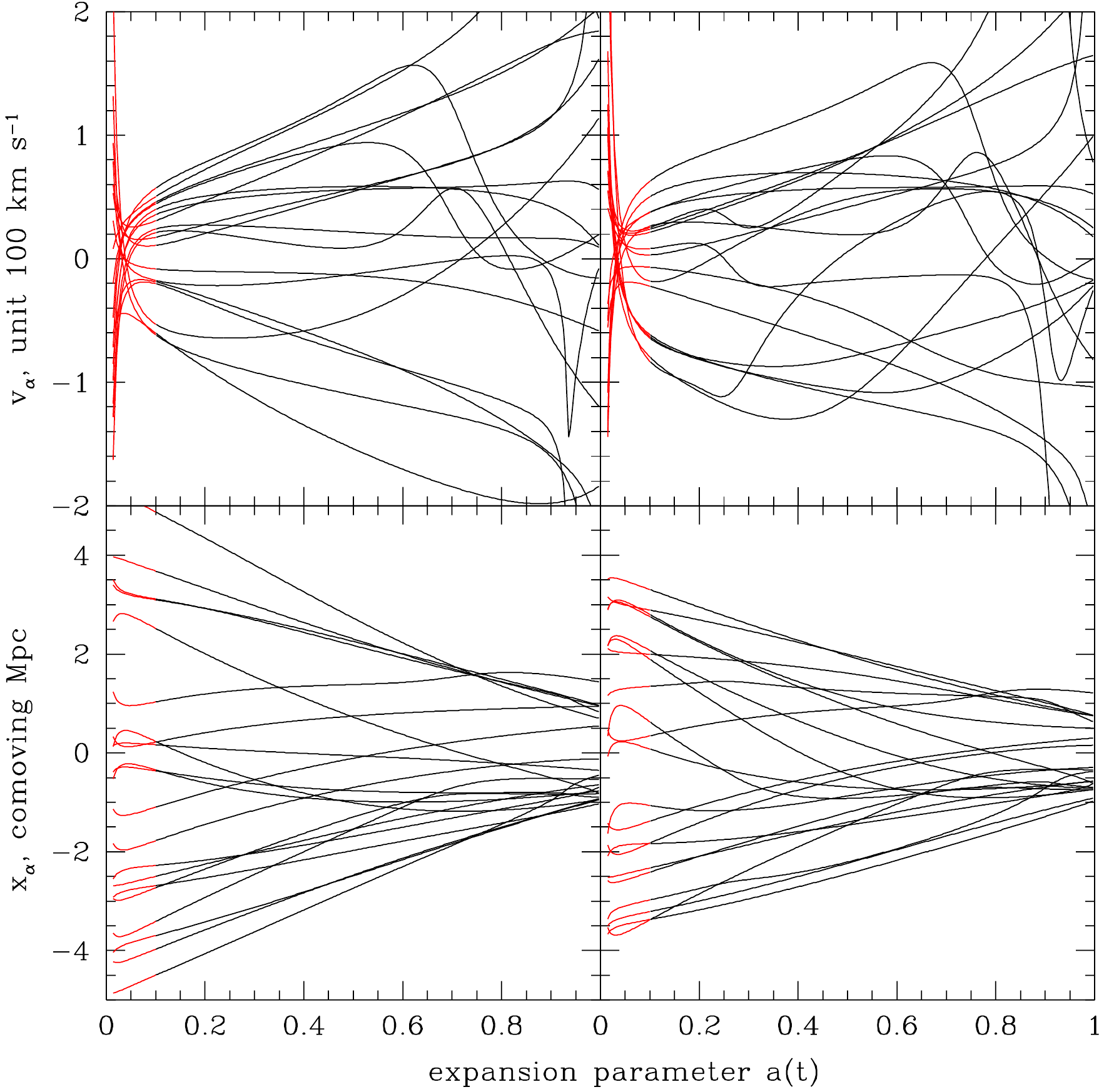} 
\caption{\label{Fig1} \footnotesize  Examples of evolution of the components of comoving positions and physical peculiar velocities. Black curves show the behavior over the time span of the NAM computation, red curves the result of numerical integration back in time from the start of the NAM computation.}
\end{center}
\end{figure}

Figure~1 illustrates the evolution of the three components of galaxy comoving positions and peculiar velocities for MW, M31, LMC, LeoI, LeoT, NGC\,6822 (N6822 in Table~1), and Phoenix (Phx). Two  solution are shown; the one on the left panel has the model parameters in Table~1. The black curves are from the numerical integration forward in time, but the NAM solution with its larger time steps is very similar. The red curves are from the same numerical integration method, but computed back in time from the start of the NAM solution. The wanted close to linear variation of comoving position with $a(t)$ near $a_i=0.1$, and the growth of peculiar velocities, are usually produced by the NAM method. The red curves show that these trends continue back to redshift $z\sim 20$. Earlier than that the peculiar velocities diverge, roughly as $v\propto a(t)^{-1}$, which is the behavior of freely moving particles. The curves in Figure~1 mimic a linear perturbation theory solution that is a mix of growing and decaying modes. The near linear variations of comoving positions with $a(t)$ at $z\la 20$, and the growth of the peculiar velocities, argue that  at $z\la 20$ the analog of the growing mode dominates, as desired.  

It should be noted that these considerations do not require that the galaxies were fully assembled at $z\sim 20$. The idea is that at sufficiently early times each particle serves as a tracer of the motion of the center of the mass that is later assembled as a present-day galaxy with its dark matter halo, which then  moves as traced by the particle.

\begin{figure}[htpb]
\begin{minipage}[]{3.9in}
\begin{center}
\includegraphics[angle=0,width=4.in,clip]{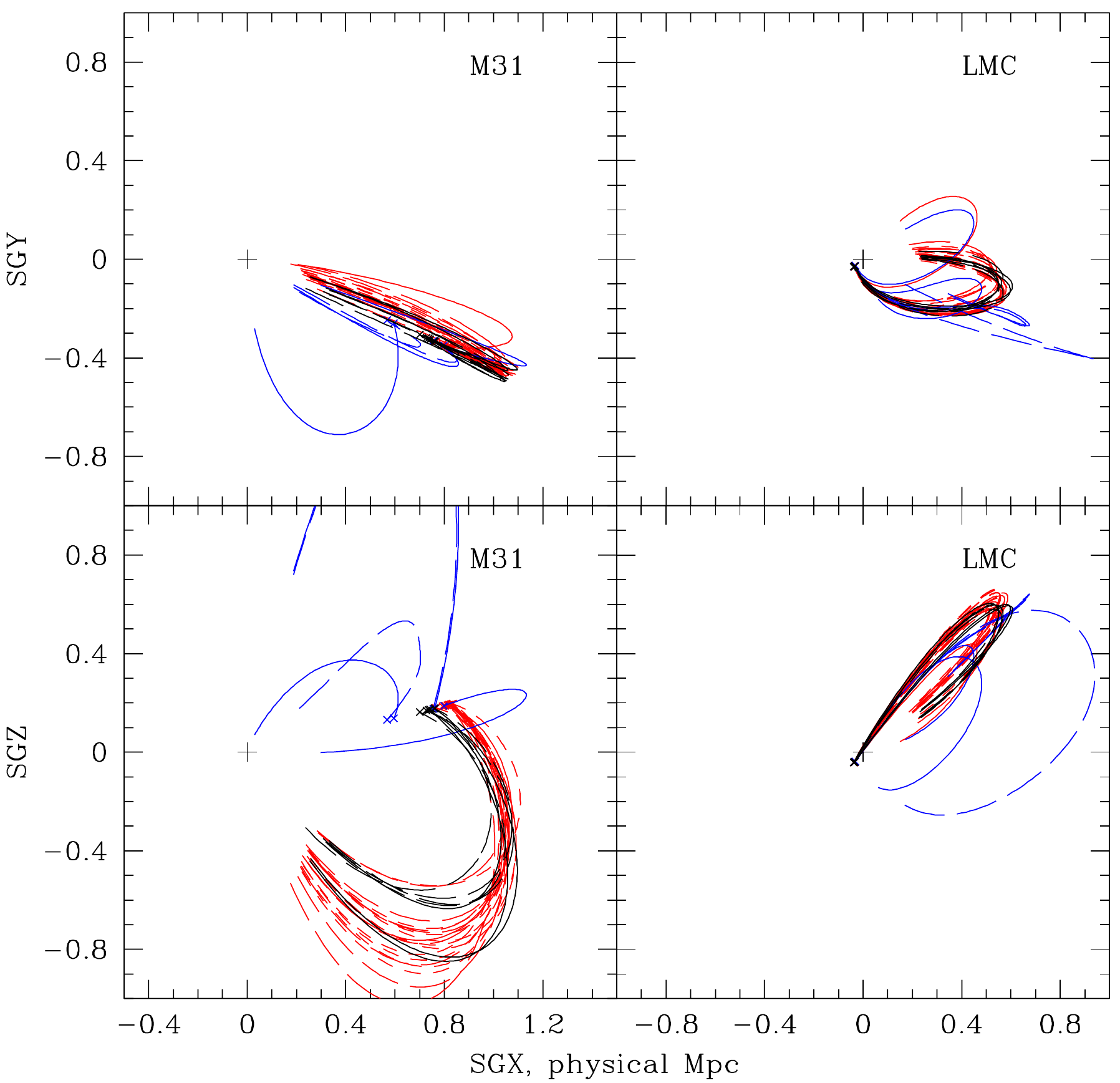}
\caption[]{\label{Fig2} \footnotesize Orbits of M31 and LMC relative to MW, in physical supergalactic coordinates. Present positions are at the crosses. The five solutions with atypical M31 orbits are plotted in blue, and the five with the most acceptable M31 distances are plotted in black. Other notation is explained in the text.}
\end{center}
\end{minipage}
\hfill
\begin{minipage}[]{2.5in}
\begin{center}
\includegraphics[angle=0,width=2.3in,clip]{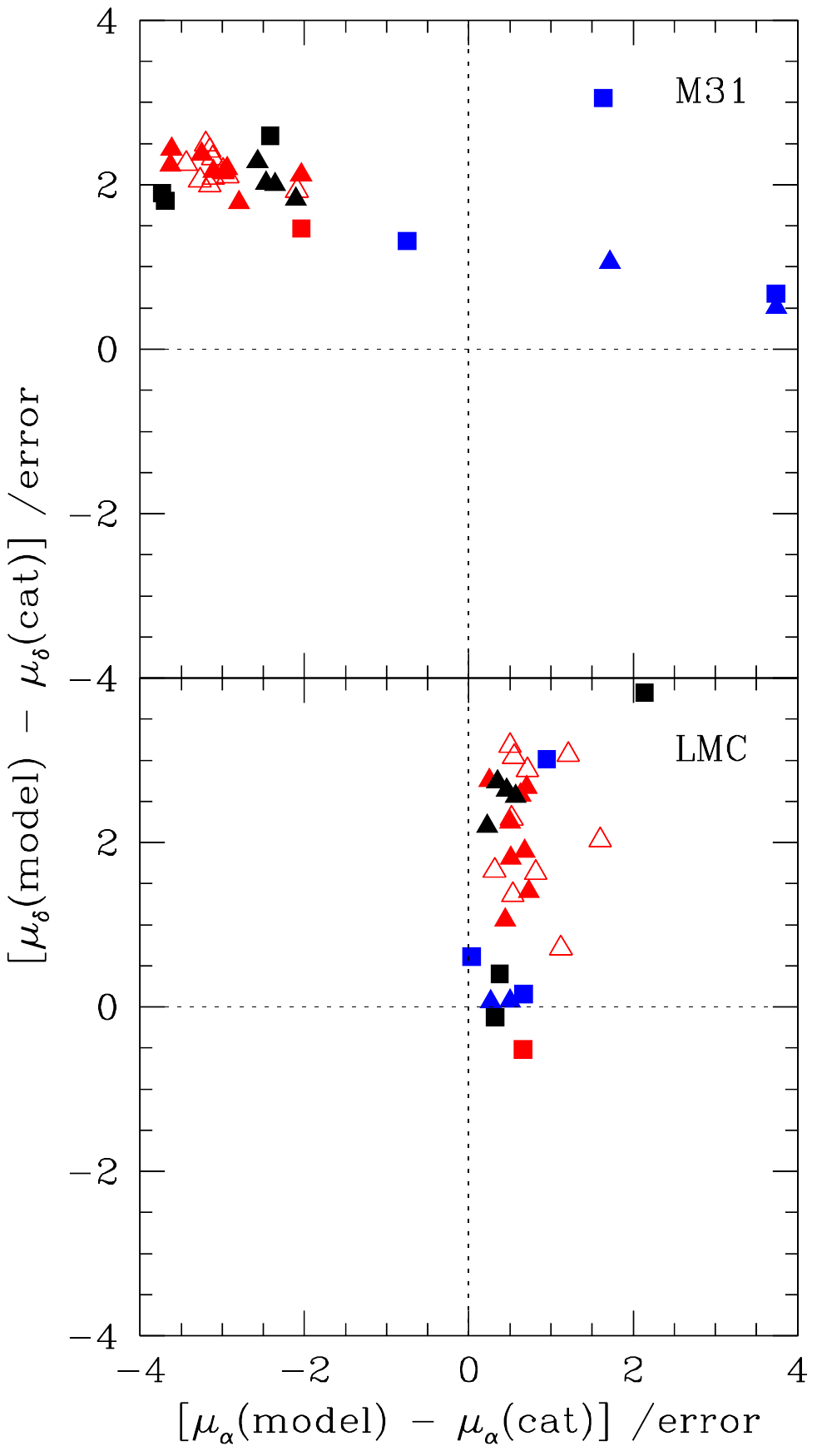}
\caption[]{\label{Fig3}\footnotesize Proper motions of M31 and LMC plotted as difference from catalog value divided by catalog uncertainty. See the text for notation.}
\end{center}
\end{minipage}
\end{figure}

\section{Solutions}

These 31 solutions allowed by the criteria in Section~2.2 are at or near different local minima of $\chi^2$. Figure~2 shows the model galactocentric orbits of M31 and LMC (that is, the positions of M31 and LMC relative to the position of MW  at the same time) in physical supergalactic coordinates. Here and in all the following figures the five solutions with M31 orbits that look quite different from all the others are plotted in blue. Among the 26 other solutions the seven with M31 distances within $2\sigma$ of the catalog value are plotted in black. The remaining 19, with M31 distances that differ from the catalog by $2$ to $4\sigma$, are plotted in red. The line type signifies the number $s_3$ of the 69 LG model parameters that differ from the catalog values by three to four times the nominal uncertainty.  The solid curves show solutions with $s_3\leq 2$, long dashes show $s_3=3$, and  short dashes show $s_3=4$ or 5. 

The heliocentric proper motions $\mu_\alpha$ in the direction of increasing right ascension and $\mu_\delta$ in the direction of increasing declination for M31 and LMC are shown in Figure~3. They are plotted as model value minus catalog value normalized by division by catalog error (Table~2). The color coding is the same as in Figure~2. Here and below proper motions in solutions with $s_3\leq 2$ are plotted as filled squares, solutions with $s_3=3$ as filled triangles, and $s_3=4$ or 5 as open triangles.  Apart from the five solutions with atypical orbits (and plotted in blue) the M31 proper motions  systematically prefer smaller $\mu_\alpha$ and larger $\mu_\delta$ than the measurement. To avoid confusion we remind the reader that these solutions are judged to be plausibly acceptable not because we have any reason to doubt the measurements, but rather because we see good reason to be cautious about the possibility of the subtle systematic errors to which dynamical models and measurements are subject. There is not much difference among the proper motions plotted in  Figure~3 in black and red with different values of $s_3$. It may be significant that some of the blue solutions with atypical M31 orbits place the proper motion of LMC close to the central value of its measurement. 

\begin{figure}[htpb]
\begin{minipage}[]{3.in}
\begin{center}
\includegraphics[angle=0,width=3.1in,clip]{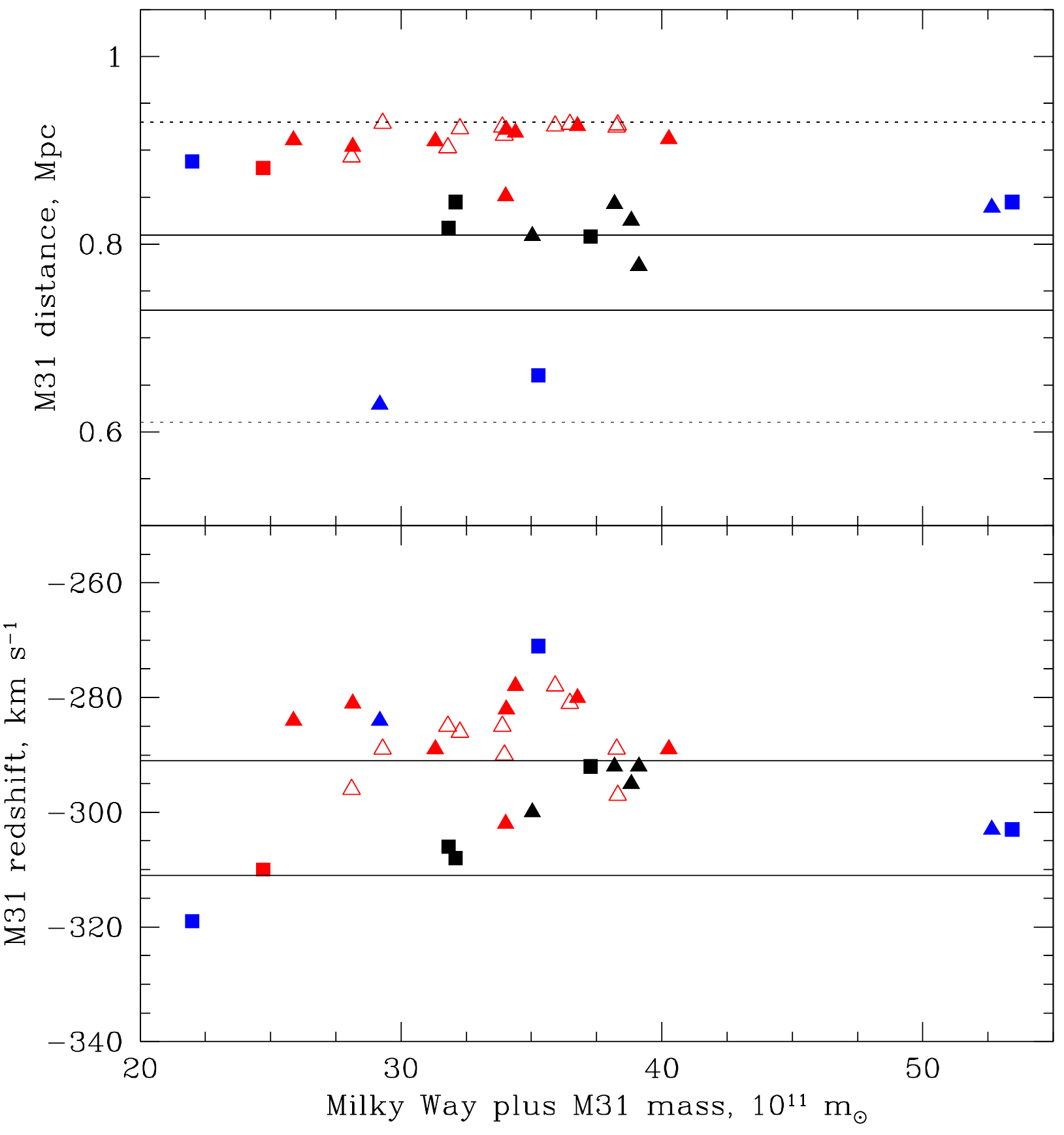}
\caption[]{\label{Fig4} \footnotesize Model M31 distances,  redshifts, and sums of MW plus M31 masses. Here and below solid lines are nominal $1\sigma$, dotted $4\sigma$}.
\end{center}
\end{minipage}
\hfill
\begin{minipage}[]{3.in}
\begin{center}
\hspace{-1.cm}
\includegraphics[angle=0,width=3.in,clip]{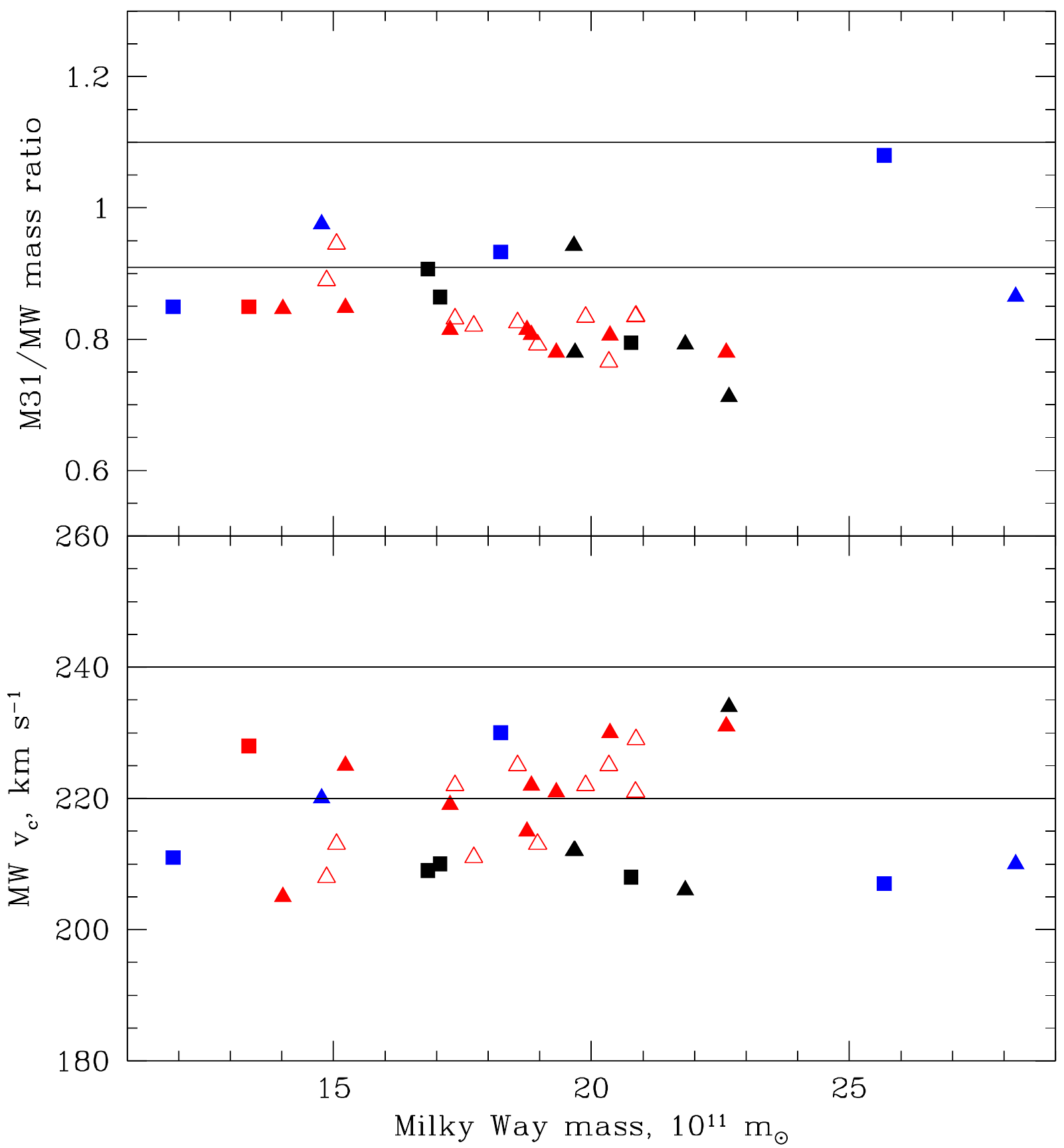}
\caption[]{\label{Fig5}\footnotesize Ratios of masses of M31 and MW,  MW circular velocity, and  MW mass.\vspace{8mm}}
\end{center}
\end{minipage}
\end{figure}

Figure~4 shows the distributions of model M31 distances, redshifts, and the sum of MW and M31  masses. The solid lines indicate 1-$\sigma$ bounds, measured for the distance and guessed for the redshift, and the dotted lines are at $4\sigma$. In many solutions with M31 distances near the allowed 4-$\sigma$ upper limit the proper motions $\mu_\delta$ of I10 are near the low end of its allowed range. Reducing the M31 distance in these solutions tends to produce an unacceptable reduction of $\mu_\delta$ for I10. In some of the plotted solutions the M31 distance was moved into the 4-$\sigma$ upper bound at the expense of moving $\mu_\delta$ of I10  toward its 4-$\sigma$ lower bound. Many other solutions were found that fit all constraints except that the M31 distance and the I10 $\mu_\delta$ both are just outside the 4-$\sigma$ bounds. The seven solutions plotted in black are selected because their model distances are within $2 \sigma$ of the catalog value. The lower panel in Figure~4 shows that the M31 redshifts in these seven solutions tend to be closer to the catalog redshift than for the rest of the solutions. 

The upper panel in Figure 5 shows the distributions of the ratio of M31 to MW masses and the MW mass. Although the $\chi^2$ measure of fit favors equal MW and M31 masses (eq.~[\ref{eq:mass-chis}]), and conventional estimates favor a larger M31 mass, M31 is less massive than MW in all solutions except the one  
with the largest sum of masses of M31 and MW, $5.3\times 10^{12}m_\odot$. One might have expected that the sum of the MW and M31 masses is more tightly constrained than the MW mass, because in the approximation of isolated radial motion the redshift, distance and expansion time fix the sum of the masses. One sees in Figure~2 that in the 31 solutions the motion of M31 relative to MW is quite far from radial, however, and in Figure~5 that the sums of the model M31 and MW masses span a considerable range. The two solutions with the largest sums of masses, 
and the solution with the smallest, 
have atypical M31 orbits. Among the rest, the sum of the MW and M31 masses ranges from 2.5 to $4\times 10^{12}m_\odot$. The range is somewhat narrower, 3 to $4\times 10^{12}m_\odot$, among the solutions plotted in black that have the best fit to the M31 catalog distance.

The bottom panel in Figure~5 shows the model MW circular velocities. Although the $\chi^2$ measure favors $v_c = 230$~km~s$^{-1}$, the models scatter around $v_c = 220$~km~s$^{-1}$, and six of the seven black points have still lower $v_c$.

\begin{figure}[htpb]
\begin{minipage}[]{3.9in}
\begin{center}
\includegraphics[angle=0,width=4.in,clip]{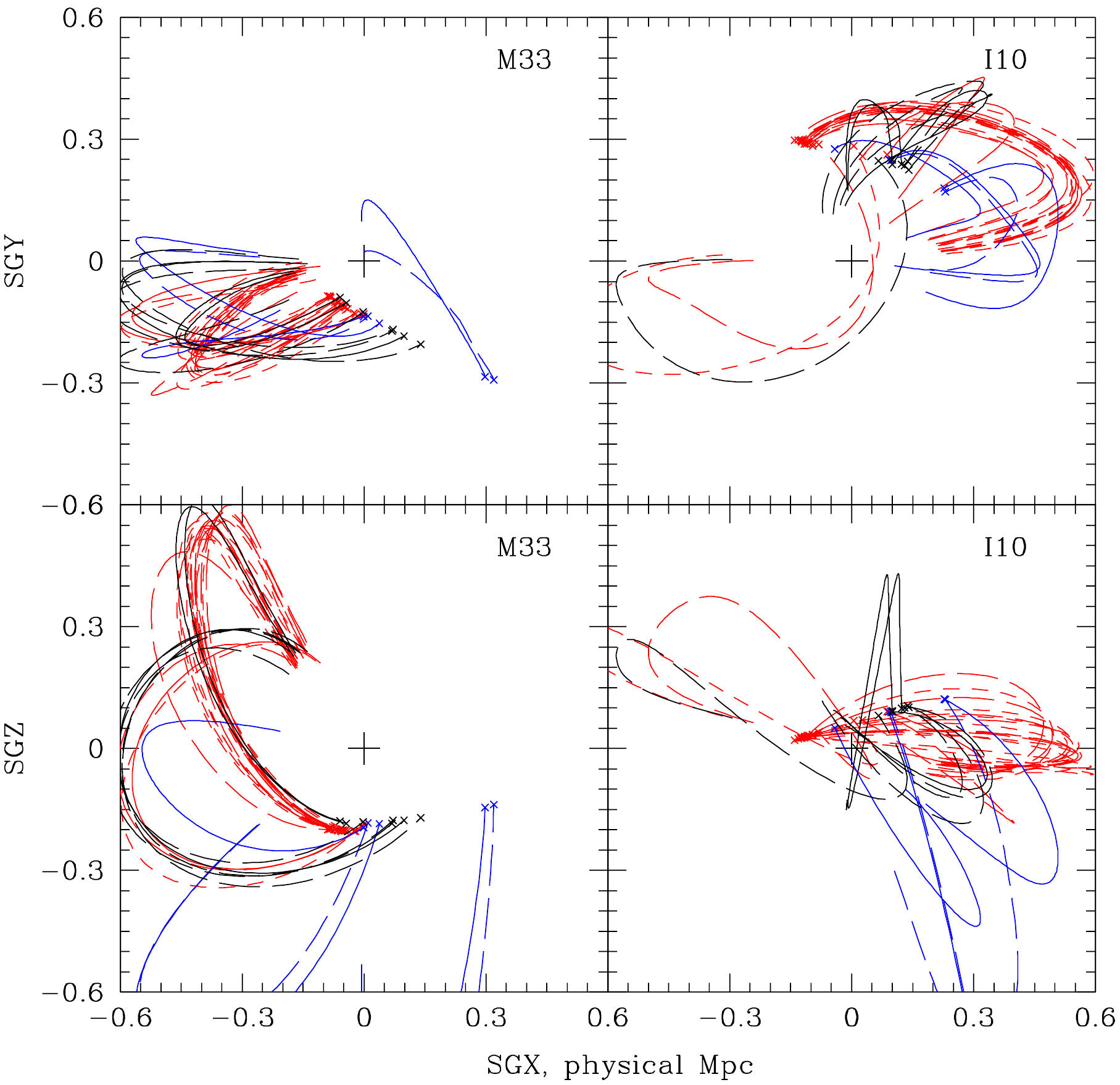}
\caption[]{\label{Fig6} \footnotesize Orbits of M33 and I10 relative to M31, in physical supergalactic coordinates, plotted as in Figure 3.}
\end{center}
\end{minipage}
\hfill
\begin{minipage}[]{2.2in}
\begin{center}
\includegraphics[angle=0,width=2.3in,clip]{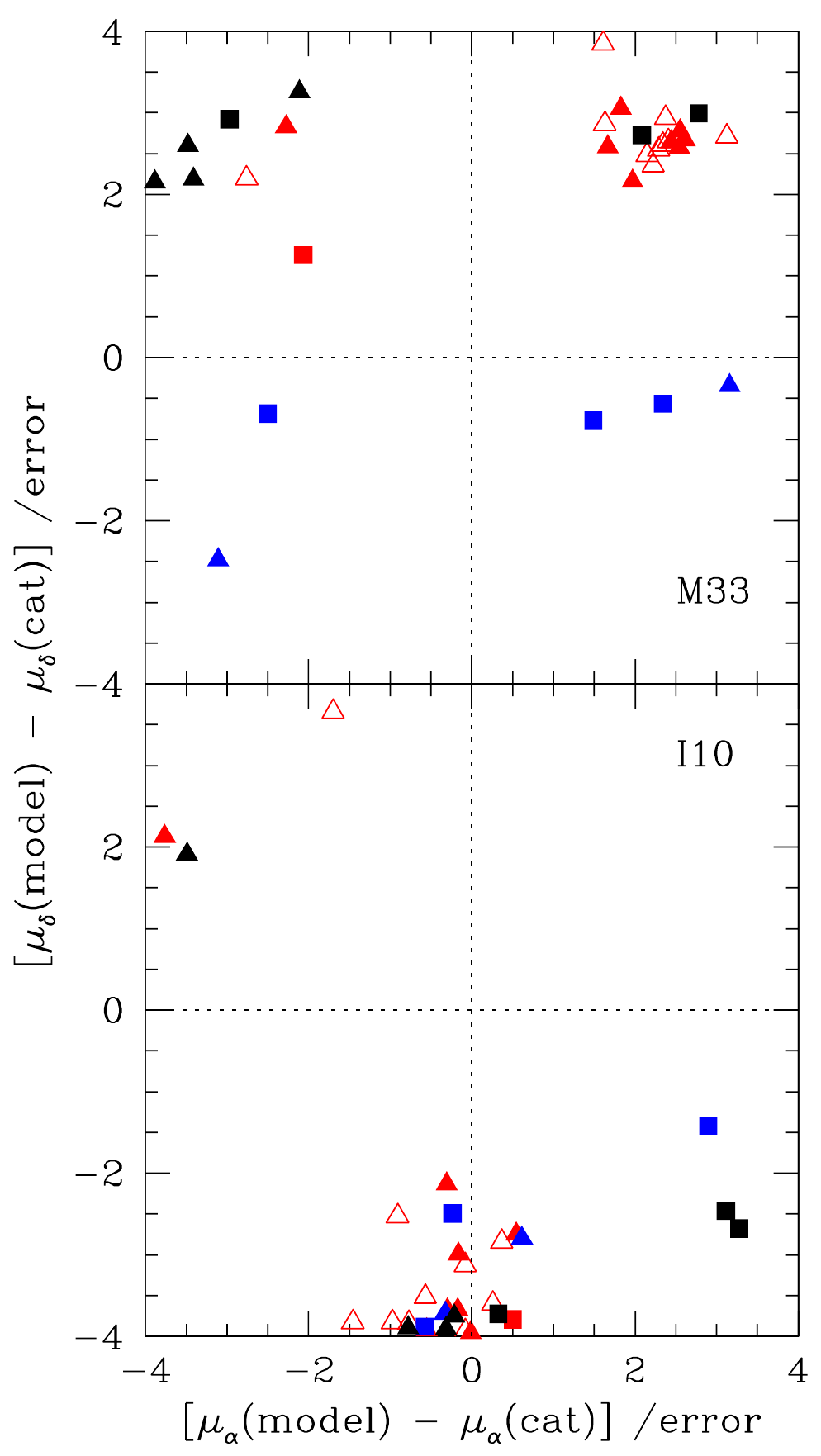}
\caption[]{\label{Fig7}\footnotesize Proper motions of M33 and I10, plotted as in Figure 3.}
\end{center}
\end{minipage}
\end{figure}

The orbits of M33 and I10 relative to M31 are shown in Figure~6, and the heliocentric proper motions of these galaxies are in Figure~7. The orbits plotted relative to MW or the LG center of mass are even more varied.  The allowed ranges of present positions, at the crosses at the ends of the curves in Figure~6, are much larger than in Figure~2  because they depend on the differences of two measured distances with similar values. The five blue solutions with atypical M31 orbits have M33 proper motions quite different from the rest, which prefer high $\mu_\delta$ and low or high  $\mu_\alpha$ relative to the measurement. The situation is even more confused for IC10. We noted that many of the solutions with M31 distance near its upper bound place $\mu_\delta$ for I10 near its lower bound. The solutions plotted in black that have the better M31 distances have no better fit to the measured proper motion of I10. 

\begin{figure}[htpb]
\begin{center}
\includegraphics[angle=0,width=4.in]{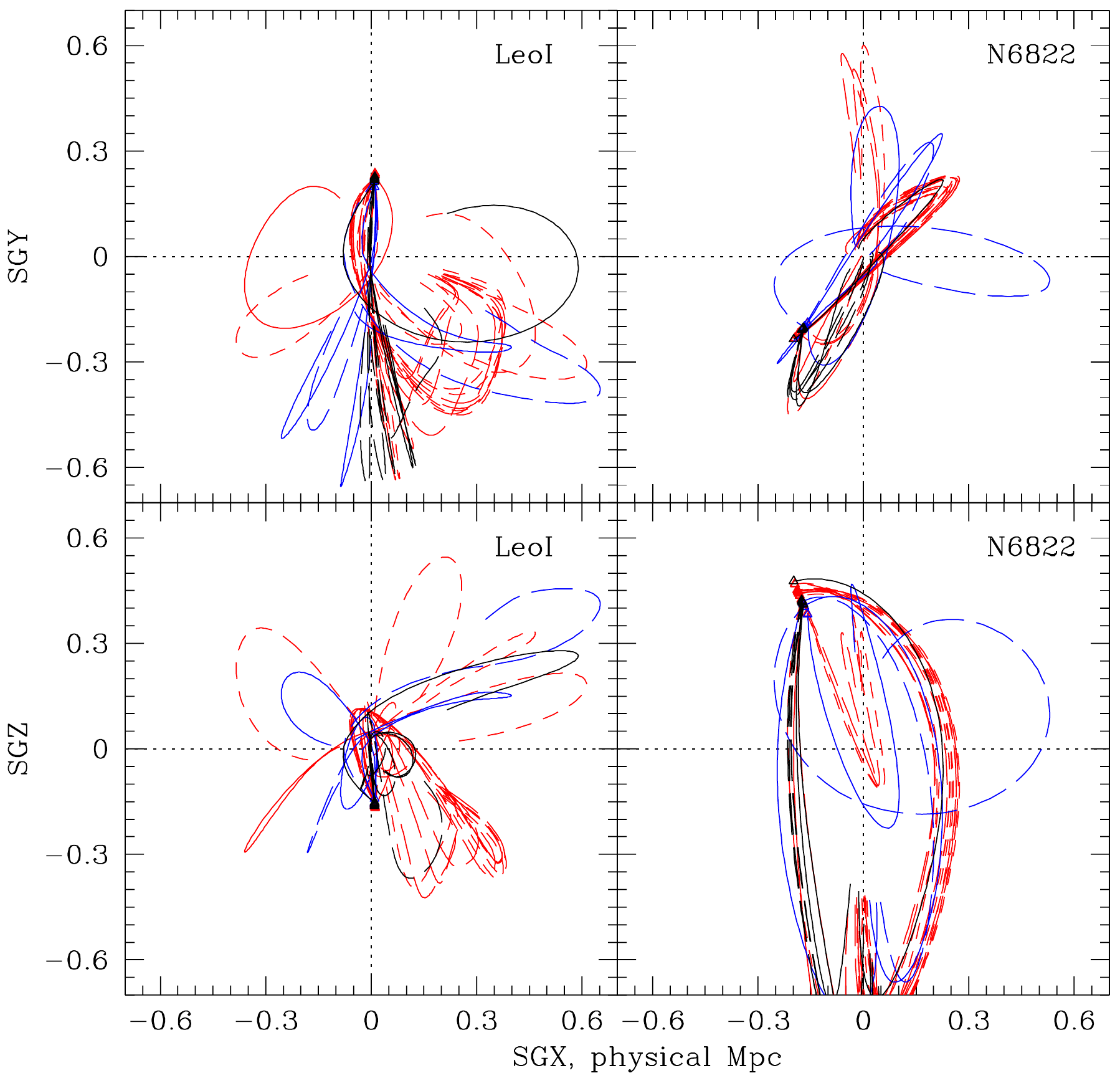}
\caption{\label{Fig8} \footnotesize Galactocentric orbits of Leo\,I and NGC\,6822.}
\end{center}
\end{figure}

Figure~8 shows galactocentric orbits of LeoI and N6822. As elsewhere some of the solutions plotted in blue that have atypical M31 orbits have atypical orbits here, while others are bunched near the rest. In all solutions LeoI has closest approach to MW in the range of 7 to 125 kpc at redshift $z\sim 0.07$, after which LeoI moves nearly radially to its present position. One can make out three types of N6822 orbits. The two that include the black curves and the solid red curves (for $s_3\leq 2$) reach physical distance from MW $\sim 700$~kpc at $z\sim 2$, and then fall back toward and past MW on one side or the other in the SGX direction. 

\begin{figure}[htpb]
\begin{minipage}[]{3.3in}
\begin{center}
\includegraphics[angle=0,width=3.3in,clip]{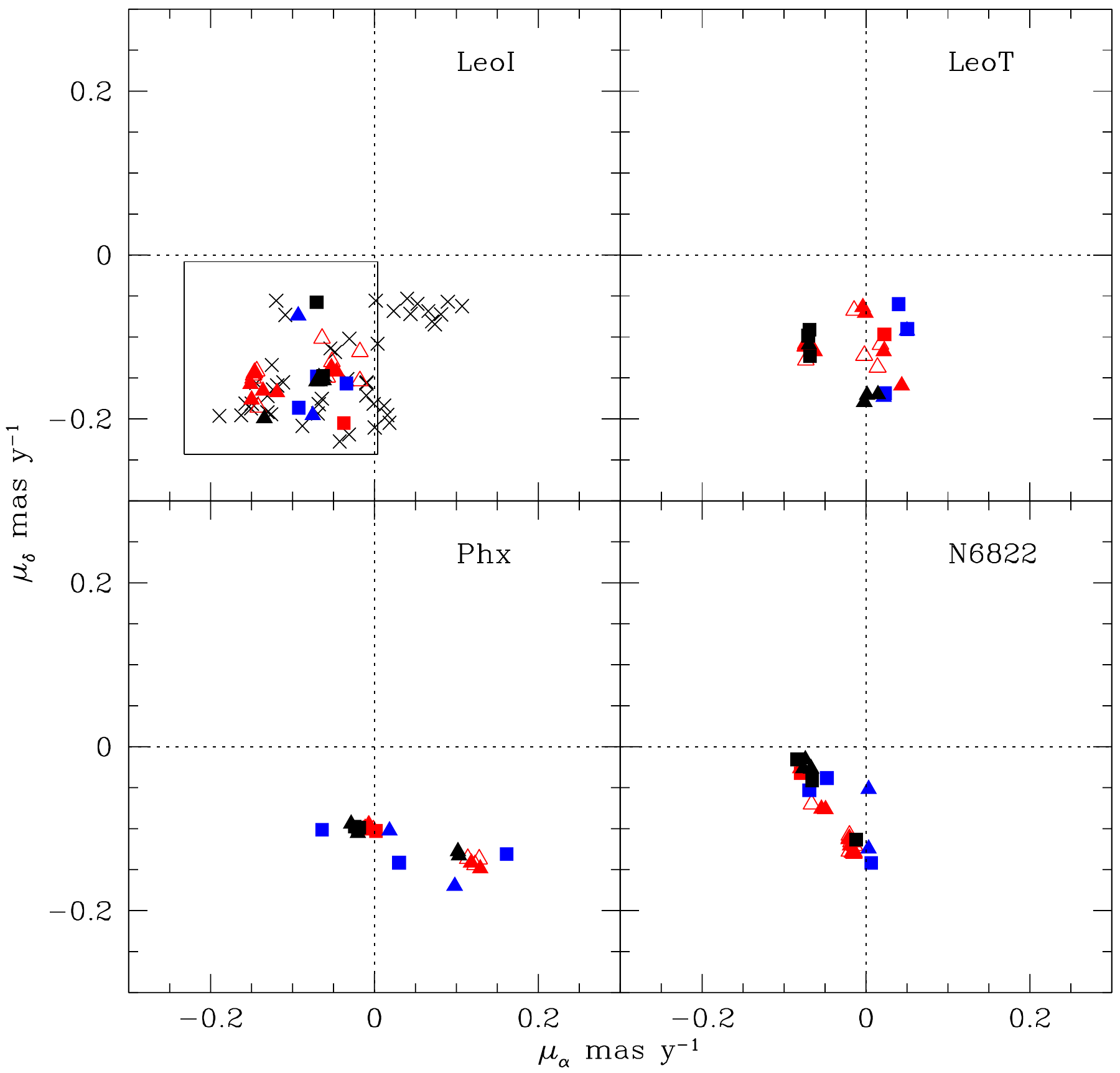}
\caption[]{\label{Fig9}\footnotesize Proper motions of LG dwarfs. The box in the upper left panel, for LeoI, is the measured 4-$\sigma$ bound. The crosses sample LeoI orbits constrained by the LeoI distance and redshift but not its measured proper motion.}
\end{center}
\end{minipage}
\hfill
\begin{minipage}[]{3.3in}
\begin{center}
\includegraphics[angle=0,width=3.3in,clip]{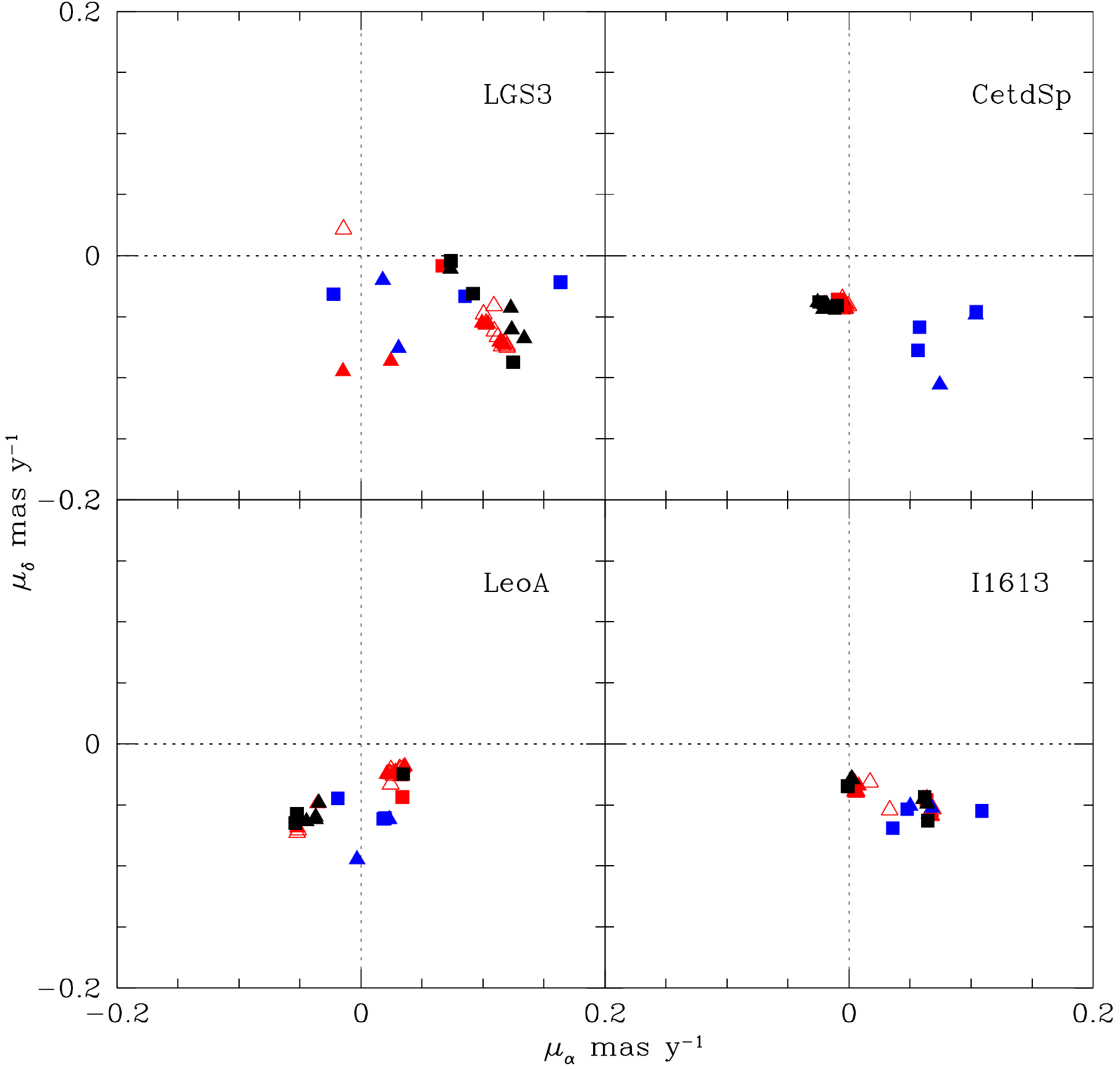}
\caption[]{\label{Fig10} \footnotesize Proper motions of more distant LG dwarfs.}
\end{center}
\end{minipage}
\end{figure}

Figure 9 shows proper motions of the closer more isolated LG galaxies, and Figure~10 shows proper motions of the more distant LG galaxies in the model. No proper motions are outside the ranges shown in the figures. The box in the upper left panel in Figure~9 shows the measured 4-$\sigma$ bounds on the proper motion of LeoI (Sohn,  Besla, van der Marel, et al., 2012). The squares and triangles are inside the box because the search for solutions required it. The crosses sample LeoI proper motions allowed by all constraints apart from this proper motion. These orbits were obtained by fixing all parameters and orbits of all actors in a solution except for LeoI, relaxing randomly chosen LeoI orbits that end at randomly chosen present distance to the equations of motion, and keeping those with LeoI distances and redshifts within the 3-$\sigma$ bounds. (This is straightforward for LeoI because its nominal mass is too small to have a significant  effect on the other galaxies. When the orbit of N6822 or one of the more massive galaxies is recast it tends to disturb other orbits, producing a noticeably different solution or even a situation that cannot be relaxed to a solution in a reasonable time by the method used here.) The crosses show two examples derived from each of the solutions (apart from a few cases where finding acceptable orbits is particularly time-consuming). One might wonder whether the crosses cluster near the box because the solutions were selected to allow a fit to the LeoI proper motion. We have a check, because many of the solutions were obtained before appearance of the LeoI proper motion measurement. These solutions were selected without attention to the LeoI proper motion, and the operation that yielded the crosses was also used to select a LeoI orbit to fit the proper motion. There is the difference that in each solution all parameters are adjusted to minimize the $\chi^2$ sum that includes the LeoI proper motion, while this was not done for the crosses. Perhaps this accounts for the crowding of crosses toward the right-hand edge of the box. But we can conclude that there are not likely to be solutions consistent with all other constraints that place the LeoI proper motion well away from the distribution of  crosses.

 \begin{table}[htpb]
\centering
\begin{tabular}{lrrrrrr}
\multicolumn{7}{c}{Table 3: Counts of 3-$\sigma$ Deviations}\\
\noalign{\medskip}
\tableline\tableline\noalign{\smallskip}
  & $cz$  &  $\ D$ & $\ m$ & $v_i$ & $\mu_\alpha$ & $\mu_\delta$  \\
\tableline
 \noalign{\smallskip}
  MW     &    0 &    0 &    0 &    0 &    0 &    0 \\
  M31    &    0 &   18 &    0 &    0 &   15 &    1 \\
  LMC    &   16 &    0 &    0 &    0 &    0 &    5 \\
  M33    &    1 &    0 &    0 &    0 &    6 &    3 \\
  I10    &    0 &    2 &    0 &    0 &    4 &   19 \\
  N185   &    0 &    0 &    0 &    0 &    0 &    0 \\
  N147   &    0 &    0 &    0 &    0 &    0 &    0 \\
  N6822  &    1 &    0 &    0 &    0 &    0 &    0 \\
  LeoI   &    1 &    0 &    0 &    0 &    2 &    0 \\
  LeoT   &    0 &    0 &    0 &    0 &    0 &    0 \\
  Phx    &    0 &    0 &    0 &    0 &    0 &    0 \\
  LGS3   &    0 &    0 &    0 &    0 &    0 &    0 \\
  CetdSp &    0 &    0 &    0 &    0 &    0 &    0 \\
  LeoA   &    0 &    0 &    0 &    0 &    0 &    0 \\
  I1613  &    2 &    0 &    0 &    0 &    0 &    0 \\
  Sclp   &    1 &    0 &    0 &    0 &    0 &    0 \\
  Maff   &    0 &    0 &    1 &    0 &    0 &    0 \\
  M81    &   24 &   25 &    0 &    0 &    0 &    0 \\
  Cen    &    0 &    0 &    0 &    0 &    0 &    0 \\
\tableline
\end{tabular}
\end{table}

\begin{figure}[htpb]
\begin{center}
\includegraphics[angle=0,width=4.5in]{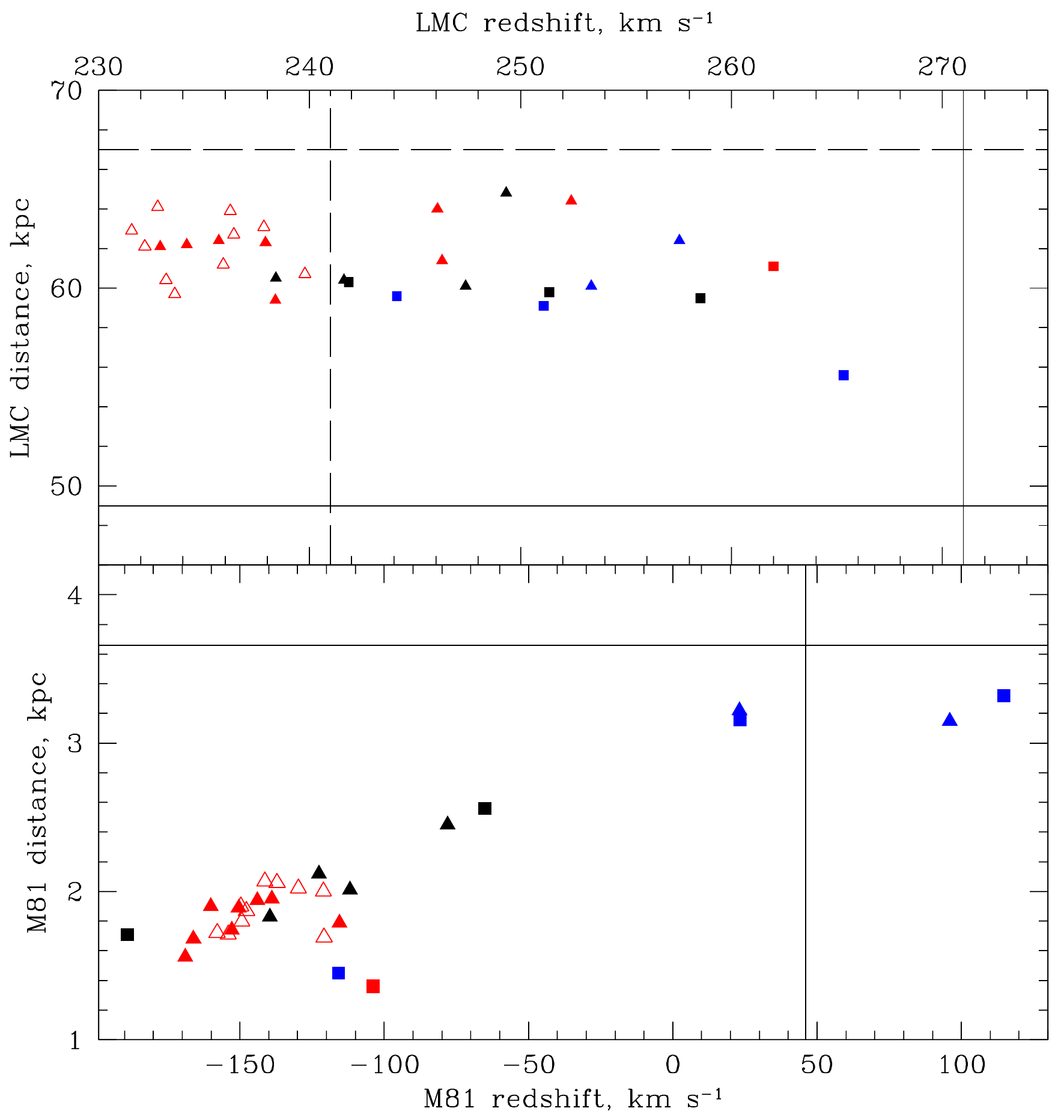}
\caption[]{\label{Fig11}\footnotesize Model redshifts and distances of LMC (top) and the external actor M81 (bottom). Solid lines are catalog central values, dashed lines nominal 3-$\sigma$ deviations.\vspace{6mm}}
\end{center}
\end{figure}

Table 3 lists the numbers of differences of model and catalog values of the redshift, distance, mass, initial peculiar velocity and proper motion that exceed the nominal 3~$\sigma$. In all 31 solutions the MW circular velocity $v_c$  and the condition on the primeval peculiar velocity are consistent with the adopted values within three times the nominal uncertainties. In one solution the mass of M31 is 0.7 times the MW mass, outside the nominal 3-$\sigma$. This solution is one of the seven plotted in black.  M31 and I10 have the most problematic proper motions (as seen in the top panel in Figure 3~and lower panel in Figure~7). M31  has the most problematic distances (top panel in Figure~4). LMC has the most problematic redshifts, though the solutions  plotted in black that better fit the M31 distance are better fits to the LMC redshift, as shown in the top panel in Figure~11. This panel also shows that the LMC distances all are larger than catalog at 2 to $3\sigma$. 

 \section{Discussion}
 
This exploration of solutions for the motions of LG galaxies allowed by a reasonably simple and well-defined mass model and numerical procedure has yielded a considerable variety of orbits at different local minima of the measure of fit to constraints. The 31 solutions include groups of orbits that have similar forms, as one sees in the three types of orbits of N6822 around MW in Figure~8. There is some pattern in the comparison of solutions with different numbers $s_3$ of the largest allowed deviation from catalog parameters. Thus in Figure~8 one of the three groups of orbits includes only solutions with $s_3=4$ or 5, and in Figure~11 one sees that the lower LMC redshifts are favored by larger $s_3$. On the other hand, the orbits and proper motions of M31 and LMC are much the same for different $s_3$ (Figs.~2 and~3). The solutions with strikingly atypical M31 orbits (Figs.~2 and~3), plotted in blue, are  enigmatic. They tend to have larger peculiar velocities at $1+z=10$, though well within the allowance. We have not found any clear reason to reject or prefer these special cases; they may signify a problem to be discovered in the dynamical model, or one of them may prove to be the most accurate approximation. The orbit of LMC relative to MW seems to be the best constrained, but even here we have only a rough indication of where LMC may have been at redshift $z=3$. But though the situation is confused  the solutions may point us to measurements and computations that could help improve the picture, as discussed in this section.

We use a semi-empirical picture of the effect of external mass that is supposed to allow the parameters of four external actors to float to approximate the gravitational field within LG produced by external mass, as judged by the fit to the LG constraints. The results of this exercise are mixed. The largest apparent anomalies are the redshift and distance of the actor M81 (Fig. 11 and Table 3). The rest of the  parameters for the external actors are within the generous assigned bounds, but it is notable that the masses of the actors Sclp and Cen tend to be about half the catalog values and the masses of the actor M81 tend to be about twice the catalog value. We suspect these anomalies are at least in part a result of compensation for an inadequate external model. We intend to explore first the simple remedy of allowing adjustment of the angular positions of the external actors (as in Peebles, Tully, \& Shaya 2011), which may offer a better picture for comparison to where the external galaxies are and how they are moving. The larger goal is to extend  explicit dynamical analyses to the more massive galaxies outside LG. 

The MW circular velocity  $v_c$ is allowed to float within generous bounds. The solutions cluster around  the conventional value, $v_c\simeq 230$~km~s$^{-1}$, despite the choice of the measure of fit to favor a larger value (eq.~[\ref{eq:vc}]). To be explored is whether our approach would be seriously challenged if $v_c$ proved to be significantly larger than 220~km~s$^{-1}$, as argued by Reid et al. (2009). The ratio of masses of MW and M31 is allowed to float within tighter bounds that favor equal masses but, again contrary to current thinking, MW is the more massive in 30 of the 31 solutions. The exception, one of the atypical M31 orbits, has one of the largest LG masses. The curious model preference for large MW mass and low $v_c$ is to be reconsidered in more complete models. 

\begin{figure}[htpb]
\begin{center}
\includegraphics[angle=0,width=3.25in,clip]{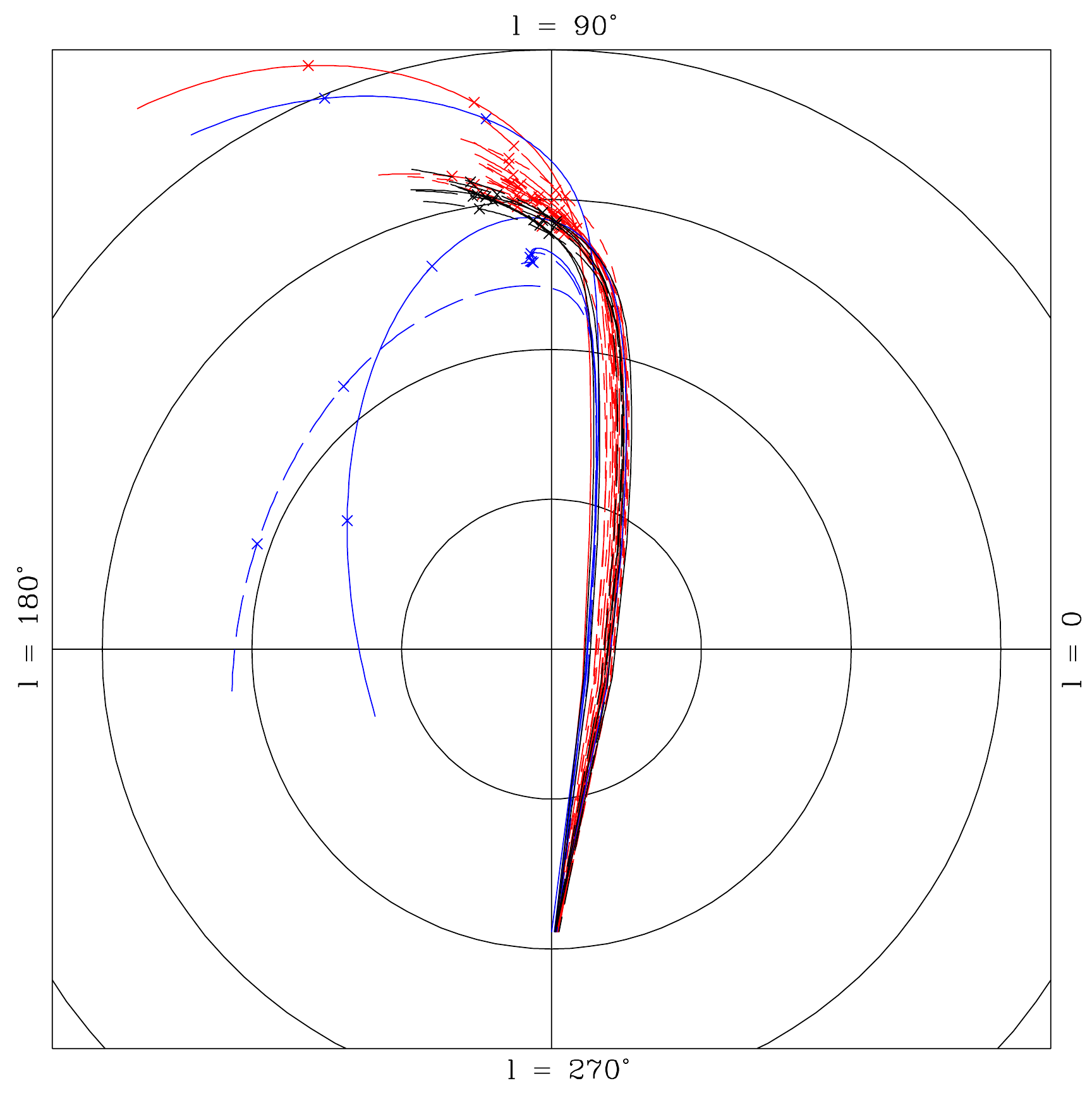}
\caption[]{\label{Fig12} \footnotesize LMC galactocentric orbits. The circles are spaced by $30^\circ$ in galactic latitude centered on the South galactic pole, the crosses mark redshifts $z=1$ and $z=3$, and the orbits commence at  $1=z=10$.}
\end{center}
\end{figure}

There is relatively little scatter among the LMC orbits in Figure~2 and the LMC proper motions in Figure~3. Figure~12 illustrates this another way, in the path of the angular position of  LMC relative to MW in galactic coordinates. Besla {\it et al.} (2007) display orbits this way, and they and Kallivayalil et al. (2013) show that the path of LMC traces back through galactic latitude $b\sim -80^\circ$ as the longitude passes through $\ell =0$. Addition of cosmological initial conditions (Peebles 2010, Fig. 3), in a minimal but possibly adequate model for a close LMC orbit around MW, indicated that the path earlier than that reaches an extremum of latitude near $\ell\sim 90^\circ$, $b\sim 0$, and at higher redshift emanates from the  direction of larger $\ell$. The relatively small scatter among orbits in this exercise suggested that the past motion of LMC relative to MW may reasonably well determined. The present model reduces the scatter in orbits, with the aid of the tighter LMC proper motion (Kallivayalil et al. 2013), but among the solutions in blue with atypical M31 orbits the scatter of LMC positions at redshift $z=3$ is similar to the scatter in the more schematic model in Peebles (2010). 
  
Van der Marel et al. (2012) conclude that the galactocentric orbit of M31 is close to radial, unlike the strongly curved M31 orbits in the solutions in Figure~2. This is in part a result of our use of the HST proper motion of M31, which is centered further from zero with larger error bars than van der Marel et al. find by using data on other LG galaxies. The HST proper motion does not exclude a near radial motion, but we find that the data on other LG galaxies force our model proper motions away from the HST central value toward still larger transverse velocities (Fig.~3). The result is that many of the 31 solutions have M31 proper motions that would be quite inconsistent with the HST measurement if systematic errors in model and measurement were negligible. We respect the great care taken in correcting the proper motion of M31 for internal motions, but the hazards of such operations in astronomy lead us to provisional acceptance of 3- to 4-$\sigma$ deviations of model from catalog here as for other parameters. Pending further information on this issue, the working conclusion from this dynamical model is that the galactocentric path of M31 is significantly curved. 

The poor fit of model and catalog distances to M31, and the preference for large distances, is a problem with the model, certainly not the M31 distance measurement. We offer four considerations. First, we noted that in many solutions the distance to M31 can be reduced toward its catalog value, but at the expense of lowering the proper motion $\mu_\delta$ of I10 below its allowed bound. This bound depends on a correction for the motion of the water maser relative to the galaxy. The correction was done with care, but a reexamination nevertheless would be welcome. Second, most of the mass of MW and M31 is in dark halos that are broadly extended enough to allow significant offsets between the observed galaxy positions and motions and the effective positions and motions of the centers of their massive halos. Dark matter halos may have tails. Perhaps there is more dark matter on the near or far side of M31 that is leading or trailing the motion of the galaxy. Perhaps the solutions with large M31 distance are telling us that the centers of the masses in MW and M31 are more widely separated than the observed centers of the galaxies. Third, the majority of our 31 solutions may simply be quite inaccurate; perhaps some parameter choice not well sampled in this study gives a much better fit to the constraints. Among the subset of seven plotted in black that place M31 within two standard deviations of the catalog distance, the M31 and LMC redshifts tend to be closer to the catalog values (Figs. 4 and 11), and the scatter in masses of MW and M31 is smaller than among all solutions (Figs. 4 and 5). Van der Marel et al. (2012) find that the sum of the M31 and MW masses is $3.17\pm 0.47\times 10^{12}m_\odot$. The scatter among all our masses is much larger, but it may be significant that among the seven solutions the scatter of masses is similar to van der Marel et al. (2012), with roughly similar central central values. One might take this as evidence that these seven solutions are more accurate than the rest, although the seven do not offer substantially better fits to other constraints. Fourth, the problem may with the crude representation of external mass. It would be curious if the external mass affected M31 or MW more than other LG galaxies, but the issue will be considered in the exploration of better external mass models. 

The considerable variety of orbits of M33 and I10 relative to M31 is at least in part a result of the large uncertainty in the distances of these two galaxies from the nearby dominant mass around M31. We have not found a solution consistent with the argument that M33 passed close to M31, leaving bridges of stars and gas (Lewis, Braun, McConnachie, et al. 2013 and references therein). We cannot exclude the possibility that this is because our numerical method is quite inefficient at finding orbits with close passages. It may be relevant that the solutions include only two close passages, LMC, whose position and velocity require it, and LeoI, which has a simple close passage of MW at high speed. A tighter measurement of the proper motion of M33, if feasible, might help clarify the issue by agreeing with one of the two concentrations of proper motion in the top panel of Figure~7, or perhaps ruling out both. 

The proper motion of I10 (bottom panel in Fig. 7) is even more problematic.  On the model side, an issue to be examined is whether our methods have missed a more accurate orbit. On the observational side, we will be looking for possible indications that the subtle correction for motion of the water maser source relative to the galaxy might be reconsidered.

The closest approach of LeoI to MW is in the range 7 to 125 kpc at redshift $z \sim 0.07$, or $\sim 1$ Gyr ago. In five of the seven black solutions the range is 18 to 27 kpc,  roughy half that of LMC, and the other two are 81 and 88 kpc, roughly twice that of LMC. At closest passage LeoI is moving through MW at $\sim 300$ to 500~km~s$^{-1}$, depending on the impact parameter. This may be fast enough that there is minimal tidal damage to LeoI. A check by numerical simulation would be feasible and interesting.

The LeoI proper motions plotted as crosses in the top left panel of Figure~9 were found by applying all constraints apart from the LeoI proper motion measurement. They show that the present model can fit the  measurement but not predict it. The crosses are close to the error box, however, leaving considerable white space in the rest of the panel. This might be counted as a positive indication for the model. The proper motions of the other galaxies in Figure~9 scatter over ranges  comparable to the uncertainty in the LeoI measurement, and the scatter in the more distant galaxies in Figure~10 is not much smaller. Significant tests of these proper motions thus seem to be feasible with current techniques, and certainly would be valuable.

The variety of solutions allowed by the mixed boundary conditions in this approach to LG dynamics has been greatly constrained by advances in the measurements. We are still left with an abundance of solutions, however, and the considerable list of open issues and ambiguities summarized in this section. Work in progress on theory and observations promises to address these issues, and perhaps improve the dynamical model or falsify it.

\end{document}